\documentstyle[twocolumn,aps,epsfig]{revtex}

\newcommand{\ie}{{\it i.e.}}
\newcommand{\etal}{{\it et al.}}

\begin{document}

\twocolumn[\hsize\textwidth\columnwidth\hsize\csname %
@twocolumnfalse\endcsname

\draft
\title{Theory of Percolative Conduction in Polycrystalline 
High-temperature Superconductors}
\author{Robert Haslinger and Robert Joynt}
\address{Department of Physics and Applied Superconductivity Center\\
University of Wisconsin-Madison \\
1150 University Avenue \\
Madison, WI 53706 \\}
\date{\today}
\maketitle

\begin{abstract}

Conduction in bulk polycrystalline high-T$_c$ superconductors 
with relatively high critical currents has been
shown to be percolative.  This phenomenon is due to weak 
links at grain boundaries.  These weak links are the major
limiting factor for technological applications which require high current
densities.  We formulate a model of these materials which 
can be reduced to a nonlinear resistor network.  The model is solved by 
analytical approximations and a new numerical technique.  The numerical
technique is variational, which makes it capable of solving a wide
variety of nonlinear problems.  The results 
show that the presence of a distribution of critical currents 
in the sample does not erase all information about the dissipative
electrical properties of individual boundaries.  
This means that an unambiguous connection can be made
between the $I-V$ characteristics 
at the microscopic level and the macroscopic electrical 
properties.  

\end{abstract}

\vspace*{\baselineskip}

]

\section{Introduction}

The discovery of high-T$_c$ superconductivity raised hopes of an important
role for these materials in applications requiring high current densities,
such as high-field magnets and transmission lines.  
The most important factor limiting current 
densities in present-day low-T$_c$ materials is flux motion, and the 
solution is to maximize pinning.  While this is important in the 
high-T$_c$ materials as well, an even more vexing problem is that of 
weak links.  Practical materials are polycrystalline: grain boundaries 
and other extended defects are unavoidable.  These boundaries, usually not 
important in low-T$_c$ materials, limit the flow of supercurrent in 
high-T$_c$ systems.  While bulk samples may be superconducting at low current 
densities, their critical currents are low.  Since a transport current 
must pass through many of these boundaries, they pose a crucial problem 
in conductor development.

Experimentally, early work on grain boundaries showed that their critical currents $I_c$ could 
vary over orders of magnitude \cite{dimos}.  Under these conditions, it 
might be expected that supercurrent flow in polycrystalline samples, 
at least near the critical 
current, would be percolative in nature.  Recent work using 
magneto-optic methods has confirmed this.  Particularly striking is work
on BSCCO/Ag tapes \cite{polyanskii}.  
As a result, one must 
understand the behavior of current flow when a {\it distribution} of 
boundaries is present, with possibly a wide spread in $I_c$.  

Onthe theoretical side, percolation theory has dealt mainly with linear circuit 
elements and focused on the critical behavior (\ie , power laws) near the 
percolation threshold \cite{ziman}.  
Some work has been done with nonlinear elements.
An example is the case of a 
lattice of resistive elements with arbitrary nonlinearity \cite{ref1}, 
\cite{ref2} where individual resistive elements are removed at
random.  The I-V characteristic
can be shown via renormalization group theory  
to follow a  power law  behavior near
the percolation threshold.  
These renormalization group calculations do not tell us anything about the 
value of the critical current, or the shape of the $I-V$ characteristic
away from the critical region.  Some work away from threshold has
been done for metal-insulator problems \cite{ref2a}.  This problem is 
in a certain sense dual to the superconductor-metal problem
studied in this paper, but the exact relation is not clear.
The case of a linear medium containing a small
admixture of nonlinear elements  has also been studied.  
This model has applications 
in the study of composites formed of  nonlinear impurities
embedded in a linear host \cite{ref3}, \cite{ref4}, \cite{ref5}.  
Theoretical predictions of $I-V$ characteristics
are made using the Clausius-Mossotti approximation \cite{ref3}, 
or the effective medium approximation \cite{ref4}.
These are limited to the case of weak nonlinearity or a low density of
nonlinear impurities.  
          
The critical phenomena approach and the weak nonlinearity case
are not closely relevant to our goal here, which is to understand the entire
$I-V$ characteristic of a strongly nonlinear system.
More closely related is the work 
of Leath and Tang.  These authors first considered a Ginzburg-Landau 
model of conduction in polycrystalline superconductors, 
and later a simpler nonlinear resistor 
network \cite{leath}.  The latter have generally been the more popular 
since then, since they provide a reasonable and relatively simple 
representation of the physical properties of a superconductor with weak links.
Hinrichsen {\etal} considered a model 
which is a special case of the ones we shall treat
\cite{hinrichsen}, and it will be discussed further below.
Our goal is different from both these papers, however. It lies in determining the 
connection of the "microscopic" distribution of link strengths to the macroscopic 
$I-V$ characteristic.  That is, we are interested not in a particular 
underlying model, but rather in determining what that model
is, given global information.  Hence we look for solution methods
which will apply to whole classes of models, and which therefore 
amount to generalizations of those which have been discovered to date.

\section{The Model}

The current flows through a random array of microcrystallites separated 
by grain boundaries which act as weak links.  In this paper, we shall 
restrict our attention to the two-dimensional case - a granular 
superconducting film.  There is nothing which prevents the 
generalization of any of the arguments to higher dimensions.
The computational method we use is independent of dimensionality.

The first step is to construct an equivalent resistor network.  
For the purposes of simplicity we shall assume that each grain is 
perfectly conducting.  This allows one to separate, at least 
temporarily, the two current-limiting factors (flux pinning and weak 
links) mentioned above.  It means that our results are limited to cases 
in which the total magnetic field (applied plus self) is small enough that 
there is no flux penetrating into the grains.  Flux only penetrates the 
boundaries.    

\begin{figure}
\begin{center}
\leavevmode
\epsfxsize \columnwidth
\epsffile{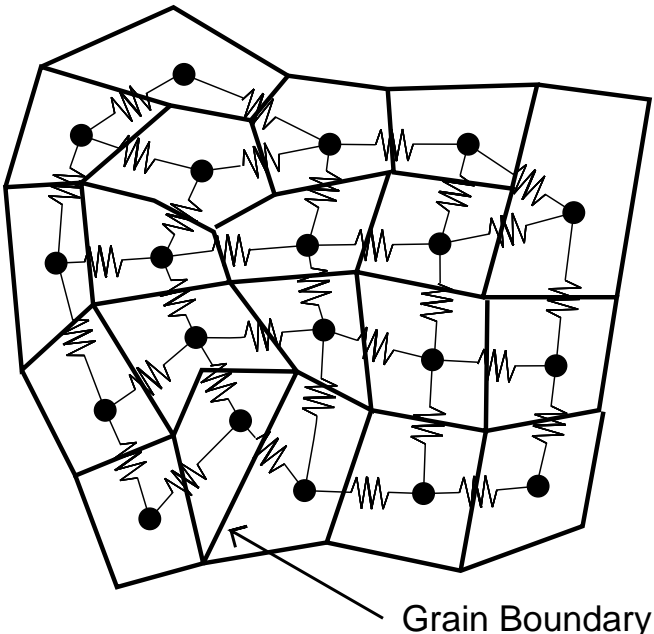}
\caption[]{Reduction of a polycrystalline superconducting
network to an equivalent resistor network.  The irregularly shaped regions
are superconducting grains, taken as equipotentials.  All potential drops 
take place across grain boundaries, shown as (nonlinear) resistors.}
\label{fig:net}
\end{center}
\end{figure}

Once the grains are taken as perfectly conducting, then they are
equipotential areas.  They are the nodes of the 
network.  Each pair of grains is separated by a boundary.  Each boundary 
is replaced by a (nonlinear) resistor.  An example of this construction 
is shown in Fig.\ \ref{fig:net}.  In 
graph-theoretical terms, the graph of boundaries is replaced by its dual 
graph.  A theorem from topology \cite{graph}
states that this relation is one-to-one.  We shall 
suppose that the geometrical randomness is less important than the 
randomness in the strength of the resistors.  Accordingly, we shall
consider only square lattice networks.

Although we will develop techniques applicable
to resistive elements of arbitrary nonlinearity, our main interest
in this paper is the flow of current through a granular superconductor.
Accordingly, we focus in this paper on two models of the grain boundaries 
applicable to the cases of Josephson junctions in
low and high applied fields.  Why actual boundaries should
behave in these ways is discussed by Likharev \cite{likharev}.  
\begin{figure}
\begin{center}
\leavevmode
\epsfxsize \columnwidth
\epsffile{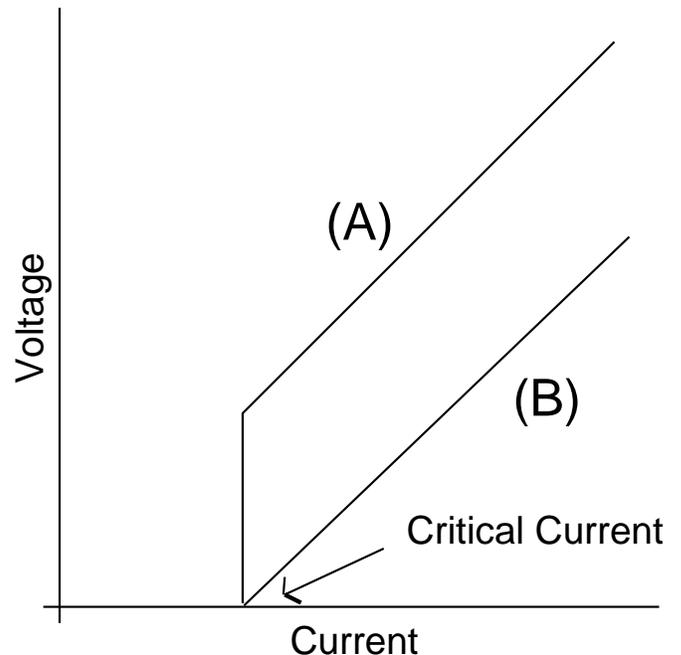}
\caption[]{$I-V$ characteristics for (A) Josephson junction (JJ) and (B) 
flux flow (FF) 
models.  The former is considered to be appropriate for low fields, 
and the latter for high fields.  In the numerical calculations, we give
the JJ characteristic a finite slope just above the critical current
for numerical stability.}
\label{fig:JJFF}
\end{center}
\end{figure}
Each boundary is specified by its
$I-V$ characteristic.  $V$ 
is some nonlinear function of $I$ for each resistor, but each resistor may
be different.  In the general case, the resistors are drawn from
a probability distribution ${\cal P}[V(I)]$.  We will
concentrate on binary distributions, since the results easiest to 
understand.  These binary distributions 
consist of a mixture of weak links and strong links, each weak link 
being identical, and each strong link being identical.
The concentration of weak links will be denoted by $p$
and the concentration of strong links by $q=1-p$.

We further divide our model into two classes corresponding to 
weak-field Josephson junction (JJ) and 
strong-field flux flow (FF) boundaries.  
For the JJ case, the weak links, with concentration $p$, 
have a critical current $I_{c1}$.  If
$I < I_{c1}$ in such a resistor, then $V=0$.  If $I > I_{c1}$, then $V = IR_1$,
where $R_1$ is a constant.  Similarly, the strong links, with concentration
$1-p$, satisfy:
if $I < I_{c2}$, then $V=0$; if $I > I_{c2}$, then $V = IR_2$.
$I_{c2} > I_{c1}$.  Note that $V$ is a {\it discontinuous} function of $I$ for 
both types of links.  Let us call this the binary JJ model.
For the FF case, the weak links, with concentration $p$, 
have a critical current $I_{c1}$.  For
$I < I_{c1}, V=0$.  For $I > I_{c1}$,  $V = (I-I_{c1})R_1$,
where $R_1$ is a constant.  The strong links, with a concentration
$1-p$, satisfy:
if $I < I_{c2}$, then $V=0$; if $I > I_{c2}$, then $V = (I-I_{c2})R_2$.
$I_{c2} > I_{c1}$.  Note that $V$ is a {\it continuous} function of $I$ for 
both types of links.  This is the binary FF model.  See Fig.\ \ref{fig:JJFF}
for an illustration of the two cases.

\begin{figure}
\begin{center}
\leavevmode
\epsfxsize \columnwidth
\epsffile{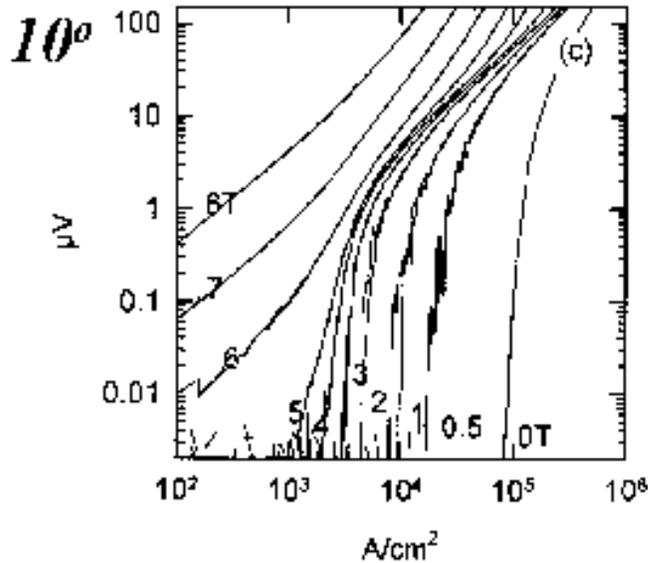}
\caption[]{Experimentally measured $I-V$ characteristics for a 10 degree YBCO grain
 boundary at various applied fields.  At low fields, the $I-V$
 characteristic resembles our JJ model.  As the applied field increases, there is
 a crossover at 5.5 T to a behavior which is similar to our FF model.  Data
 courtesy of N. Heinig and D.C. Larbalestier.}
\label{fig:tendeg}
\end{center}
\end{figure}

The two models represent very different microscopic pictures.  The JJ 
model represents a low-current, zero applied field situation.  The 
junction is taken as the simplest sort of Josephson junction.  It is 
superconducting for small currents.  It is normal at high currents and 
the voltage drop is just ordinary Ohmic loss.  There is a 
discontinuous change between the two regimes.  
The FF model is intended to simulate a situation in which the links are 
weak because flux pinning in them is weak.  In this case, $I_c$ of the 
link represents the depinning current.  The losses for $I>I_c$ are from 
the movement of flux along the boundary.    
Experimental measurements on individual grain boundaries  
show a cross-over between the two sorts of behaviour.  
Fig.\ \ref{fig:tendeg} shows $I-V$
characteristics for a 10 
degree grain boundary.  At low applied
field, the boundary resembles our JJ model, while at high field 
it is closer to the FF model.

As to the more fundamental issue of whether our model
is really valid for superconductors, we note that 
the main missing element is the phases of the grains.
This may not be as bad an approximation as at first sight, however.
Our nonlinear network can be considered as an approximation
to a true Josephson junction array in the overdamped limit.  
The JJ model can represent a set of overdamped junctions with no 
pinning in which the McCumber
parameter $\beta_J \equiv \hbar R^2/(2eI_{c0}C)^{1/2}$ is less than 1.
$R$ and $C$ are the resistance and capacitance of the junctions.  
Our output $I_{tot}$ then represents the 
DC component of the actual output of real overdamped junctions.  
The FF model is the case where pinning of Josephson vortices
dominates the transport.  

\begin{figure}
\begin{center}
\leavevmode
\epsfxsize \columnwidth
\epsffile{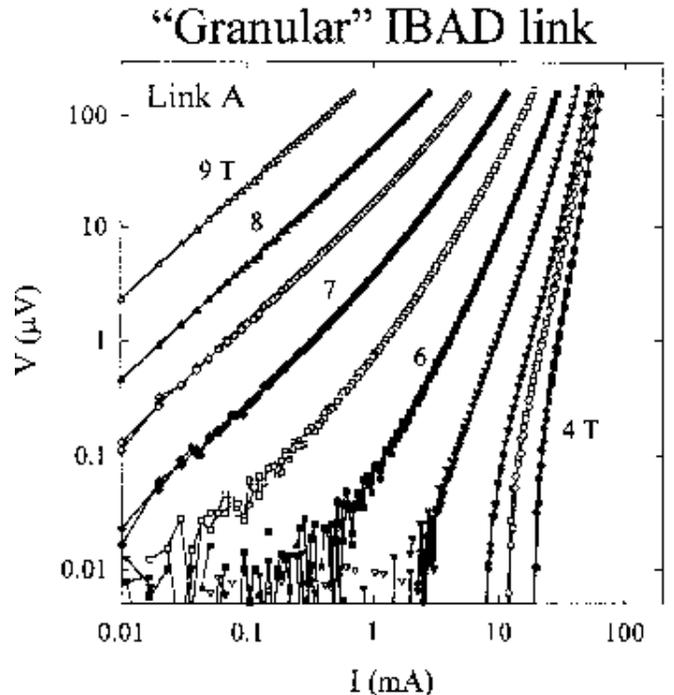}
\caption[]{Experimentally measured $I-V$ characteristics for IBAD film
 at various applied fields.  Again there is a crossover
 at roughly 5.5 T.  As in our model (Fig. 9) the {\it shape} of the individual
 grain boundaries $I-V$ is preserved in the macrosopic system.  Data
 courtesy of N. Heinig and D.C. Larbalestier.}
\label{fig:ibad}
\end{center}
\end{figure}

The output of our calculations will be a $V_{tot}(I_{tot})$ relation 
for the system as a whole.  Thus, we feed a fixed current in at 
one end and collect it at the other, and measure the voltage drop.  
We wish to compare this result to experimentally measured $I-V$ 
characteristics such as that shown in Fig.\ \ref{fig:ibad} and determine the extent 
to which the percolation is responsible for the macroscopic 
electrical properties and in particular, the critical current.
In particular, we wish to understand whether the presence of randomness
and nonlinearity washes out most of the information about the individual
resistors.

\section{Analytical Approximations}

In this section, we present an analytic approximation for the
critical current of a binary lattice.
Suppose we have a distribution of nonlinear resistors on an infinite lattice
with each resistor possessing some critical current and occupation
probability. (We shall take a lattice
constant of unity to avoid distinguishing between currents and
current densities.)  This implies that the critical current can be found from
the prescription:
\begin{equation}
I_{ctot} = min_{S} \sum_{\ell \in S} I_c^{(\ell)},
\label{eq:rhyner}
\end{equation}
where we take the minimum over {\it all} surfaces 
which separate the electrodes \cite{rhyner}.  $\ell$ are the wires which pierce
$S$, and $I_c^{(\ell)}$ is the critical current of wire $\ell$.
This explicit but highly formal equation does not give any prescription for computation
of $I_{ctot}$ except in very small systems.  It says only
that the network is only as strong as its weakest hypersurface;
in two dimensions, its weakest curve.  However, because 
$I_{ctot}$ depends only on 
$I_c^{(\ell)}$, Eq.\ \ref{eq:rhyner} tells us that
the total critical current of the lattice depends only upon
the critical currents of the individual elements and not upon
their dissipative behaviour at $I>I_c$.   

Consider two
adjacent infinite surfaces $S$ and $S'$, both of which
separate the electrodes.  The maximum supercurrent which can be 
carried through $S$, considered entirely by itself, is clearly
\begin{equation}
I_c^{(\infty)} = \sum_j p_j I_{cj}.
\label{eq:bound}
\end{equation}
However, it is not clear that the current can penetrate from
$S$ to $S'$.  Indeed we expect the critical current of the infinite lattice
to be less than $I_c^{(\infty)}$, which serves as an
upper bound for $I_{ctot}$, independent of model and dimensionality.
 
To obtain a better approximation for $I_{ctot}$,
consider a binary model, where there are only two
types of resistors $I_{c1}$ and $I_{c2}$ with occupation probabilities
$p$ and $q=1-p$ respectively.
Since we are only interested in the critical
current, the exact $ I-V $ characteristic of these
resistors in unimportant.  For this model the bound of Eqn.\ \ref{eq:bound} is 
$I_c^{tot} = pI_{c1} + q I_{c2}$.  
As $q$ increases,
there will be a large increase in the 
the critical current when  $q > p_c$.  This is obvious since
for $q>p_c$ we get an infinite cluster of $I_{c2}$ resistors running
through the lattice, allowing much more current to be transported.
However, it is {\bf not} true that $I_{ctot}= I_{c1}$ for $q<p_c$.  
This is most easily understood in the following manner.

\begin{figure}
\begin{center}
\leavevmode
\epsfxsize \columnwidth
\epsffile{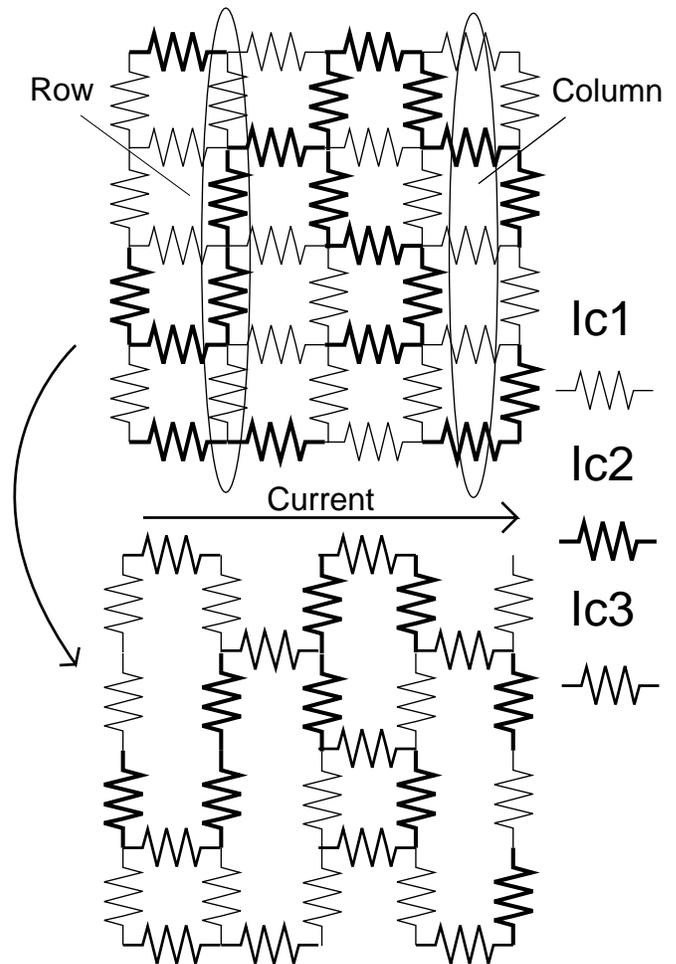}
\caption[]{Picture of a two-dimensional network.  The current runs to
the right.  We divide the resistors into those parallel to the current
(column) and perpendicular (row).  The actual lattice is then reduced
to a dilute network by subtracting out a uniform current to the right.}
\label{fig:red}
\end{center}
\end{figure}

Consider an $M \times N$ square lattice and 
suppose there is a current $I> N I_{c1}$ transported 
horizontally across the lattice.
Due to the geometry of the lattice, we 
divide the resistors into two groups.
We will call those resistors running perpendicular to the average
current flow "row resistors" and term a line of these to be a "row".
The rows are connected by what we will call "column resistors" and
the resistors connecting two rows make up a "column",as 
shown in Fig.\ \ref{fig:red}.
The current flow can be considered a superposition of two current
distributions, $NI_{c1}$ running directly through the lattice (along
the column resistors),
and a percolating current $I - NI_{c1}$.  It is this percolating
current that we are interested in, so we can "subtract off" an $I_{c1}$
resistor from each column resistor. A modified dilute 
resistor network lattice results and is shown in Fig.\ \ref{fig:red}
The columns of this lattice have holes with probability p,
and $I_{c3}=I_{c2}-I_{c1}$ critical current resistors with probability
$q=1-p$.  This modified lattice has a critical current $I_{cM}$
The total critical
current is then $I_{ctot}=I_{c1} + I_{cM}$.
$I_{cM}$ is determined by two factors.
First, the amount of current that can be transported across any 
given column 
of the lattice, and second, whether or not that current
can be redistributed to the $I_{c3}$ resistors in the next column.  There
are two different regimes of behavior distinguished by whether $I_{c3}$ is
less or greater than $I_{c1}$.  The first case is simplest.

When $I_{c3}<I_{c1}$ $(I_{c2}<2 I_{c1})$, 
it is fairly easy to redistribute the current between
columns.  Any current coming across a column resistor can be shunted 
sideways along a row  and redistributed to the next column as long as 
another column current doesn't get in the
way.  We can approximate the probability that a given column current can
be redistributed along a row without being interfered with 
in the following fashion.  It 
is equal to the probability that there is an $I_{c3}$ resistor directly
across from the given column 
current, plus the sum of probabilities that there
is a path at a site $s$ steps along the row , and no paths either coming
in or out of the row before that point.  By only summing over steps to 
the right (or left)
of the given column current, we avoid the problem of interfering with the 
paths of other incoming column currents.
This redistribution probability can then be written as:
\begin{equation}
p_{r} = q \times \sum_{s=0}^{\infty} {(1-q)}^{2s} = \frac{q}{2q-q^2}
\end{equation}

Since we have a volume fraction $q$ of $I_{c3}$ row currents, we can then
write down the averaged critical current density as:
\begin{eqnarray} 
I_{ctot} &=&  I_{c1} + (I_{c2}-I_{c1}) \times q \times p_{r}
\nonumber\\
&=& I_{c1} + (I_{c2}-I_{c1}) \times \frac{q^2}{2q-q^2}.
\label{eq:pro}
\end{eqnarray}
It should be evident that this is only an approximate formula.  The
derivation neglects the possibility that the percolation current goes 
backward, for example.  However, we expect this path to make a smaller 
contribution than the ones calculated, since they must involve 
relatively rare configurations of resistors.

\begin{figure}
\begin{center}
\leavevmode
\epsfxsize \columnwidth
\epsffile{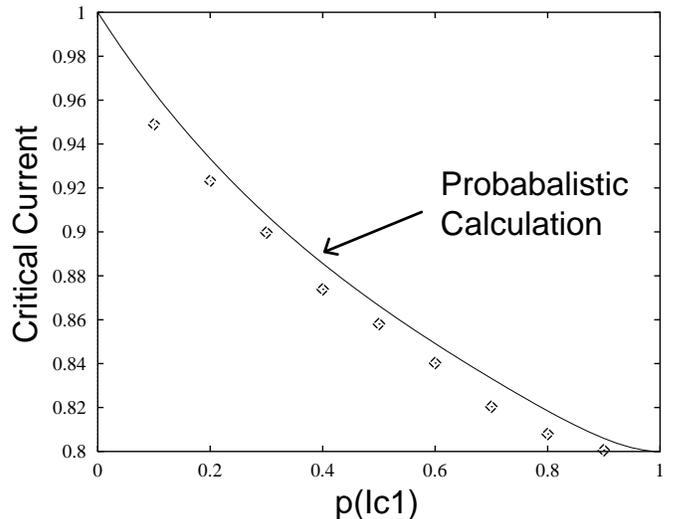}
\caption[]{Comparison of probabilistic theory to numerical data. The critical
current  for a square lattice where  $I_{c1}/I_{c2}=0.8$  is approximated 
for varying occupation probabilities p
and compared with the numerical data for a $15 \times 15$ lattice.}
\label{fig:overs}
\end{center}
\end{figure}

A plot of Eq.\ \ref{eq:pro} for a 15 by 15 lattice 
is shown in Fig.\ \ref{fig:overs}, where the results are compared
to numerical data.  It appears to 
overshoot the critical current by a substantial amount.  
We will see, however,
that this is due to the finite size of the lattice used (15 by 15). 
For the infinite lattice case, the approximation should be
very good if $I_{c3} <I_{c1}$.  

If $I_{c3}>I_{c1}$ then this 
calculation will fail, since
it is no longer as easy to shunt the current sideways 
along a row of resistors,
and hence the redistribution probability will be different.
A calculation similar to that above gives
a different critical current:
\begin{equation}
I_{ctot} = I_{c1} + q \lbrace I_{c1} (p_r - p_{r}' )
+ (I_{c2}- I_{c1}) p_{r}' \rbrace
\end{equation}
where:
\begin{equation}
p_{r}' = q \times \sum_{s=0}^{\infty} (p^2q)^{s} = 
\frac{q}{1-p^2q}
\end{equation}

While this analyisis can clearly be extended to distributions 
of more than two resistors, it quickly becomes much more complicated.
In addition, it tells us nothing about dissipative behaviour at
currents greater than $I_{ctot}$.  Hence we now turn to numerics.

\section{Numerical Calculations}

Assume that we have a nonlinear resistor network and that
the distribution of resistive elements is known.  
If a current is imposed across
the network, a voltage drop results.  We wish to calculate this 
voltage drop for the entire lattice. The correct
distribution of currents in the lattice can be found by solving
Kirchoff's equations, which uniquely determine the current distribution
in the network.  In a network of linear resistors, this reduces to solving a set of
linear equations, for which many standard methods exist.  For nonlinear 
resistors, we are not aware of any general method.  The method of 
Leath and Tang \cite{leath}, also used by Hinrichsen \etal,
\cite{hinrichsen}, works only for the JJ case.
The approach we take is to search among all distributions of current which 
satisfy current conservation and transport the imposed current across the 
lattice.  We then determine which distribution in this class also
has a unique voltage at each node.  

\begin{figure}
\begin{center}
\leavevmode
\epsfxsize \columnwidth
\epsffile{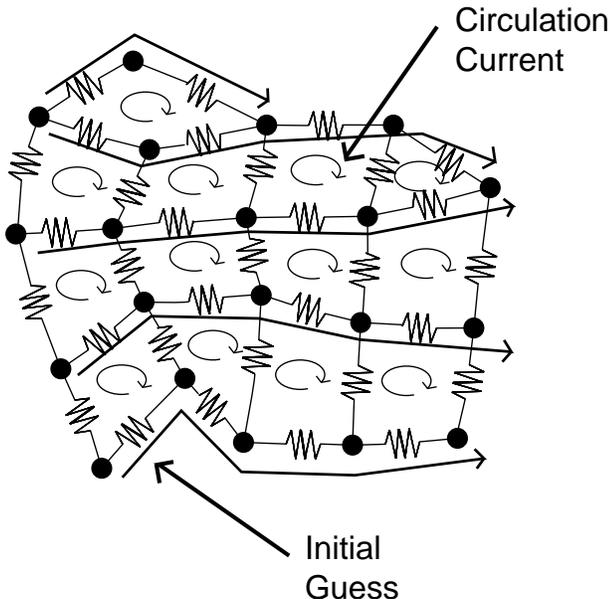}
\caption[]{Initial guess for the current distribution and imposition of
circulation currents}
\label{fig:init_imp}
\end{center}
\end{figure}

To search among all current distributions in this class, 
we begin by choosing a distribution
which is thought to be close to the actual one.  This choice 
must conserve current at each lattice node, and must also transport the 
imposed current across the lattice.  One possibility would be a distribution
where the current moves uniformly across the 
lattice.  
This initial guess 
obviously needs to be modified.  We do this by superimposing circulation currents
on top of the initial distribution as shown in Fig. \ref{fig:init_imp}. 
While these circulation currents contribute
nothing to the net transport of current, they alter the path that the 
current takes.  Current conservation is obviously satisfied for all values
of the circulation currents.  The circulation currents have become
our free variables.  It is an elementary result of homology theory 
\cite {graph} that 
all circulating currents can be generated by the addition of currents 
circulating around plaquettes.  Thus our search method can ultimately 
produce all possible currents.  

The current distribution satisfying 
the voltage Kirchoff law must still be 
found.  We note that this requirement can be 
expressed in terms of a minimization
principle.  If, for a given resistor, we define the quantity:
\begin{equation}
g_j = \int_0^{I_j} V_j (i) di
\end{equation}
where $V_j(i)$ is the voltage-current characteristic of the jth resistor (which
we assume to be known) and $I_j$ is the current flowing through the jth resistor,
then solving Kirchoff's laws is equivalent to minimizing:
\begin{equation}
{\cal G} = \sum_j g_j
\end{equation}
where the sum is over all resistors in the lattice. 
We prove this by varying ${\cal G}$ with respect to the currents  such
that current conservation is preserved.  That is, 
vary ${\cal G}$ with 
respect to a circulation current:
\begin{equation}
\frac{\delta {\cal G}}{\delta C_k} 
= \sum_j V(I_j) \frac{\delta I_j}{\delta C_k} = 0
\end{equation}
Since the circulation current $C_k$ travels around 
a specific loop, variations in
$C_k$  only affect the currents along that loop.  
In addition the effect is
unit linear in $C_k$. 
\begin{eqnarray}
\frac{\delta I_j}{\delta C_k} &= 1&  \; (I_j \> 
on \> loop \> k) \nonumber\\
\frac{\delta I_j}{\delta C_k} &= 0&  \;  (otherwise)
\end{eqnarray}  
where the sign is positive in the direction of $C_k$.  Then,
\begin{equation}
\frac{\delta {\cal G}}{\delta C_k} = \sum_{j \in loop} V_j = 0
\end{equation}

This is valid for any loop in the lattice and hence we see 
that the minimization
of $\cal G$ with respect to the circulation currents is equivalent 
to solving Kirchoff's
voltage law.  It is interesting to note that in 
the case of linear resistors, 
$\cal G$ reduces to ${\cal G} = (1/2) \sum_j R_j I_j^2 $.  
So for the linear
case, or indeed any network where all the voltages 
have the same power law dependence
upon current,  minimizing $\cal G$ is equivalent 
to minimizing the power loss of
the lattice, a familiar result.

It seems unlikely to us that this variational principle for nonlinear
systems is not known.  However, we have not been able to find it in the
literature.  We note that this method is not at all limited to this electrical example.
It should be applicable to nonlinear transport of any conserved quantity.
Fluid flow in random porous media is one candidate.  
 
Due to its robustness and applicability to highly disordered systems with 
many variables we chose a simulated annealing algorithm to perform the
minimization.  
Determining whether an optimization algorithm has found the global minimum can
be difficult.  However, in the nonlinear percolation case with monotonic 
$I-V$ relations,  a check 
can be made.  At the global minimum, the voltage drop across the lattice
should be a constant regardless of the path taken.  By calculating the voltage
drop along various paths, the accuracy of the solution can be determined. 

The main advantage of this algorithm is its generality.  It is immediately
applicable to elements of {\it arbitrary}
nonlinearity, as well as any geometrical arrangement of these elements.
We also note that this algorithm supposes an imposed {\it current} across
the lattice.   An equivalent minimization principle for an applied  voltage
can be formulated by minimizing the sum:
\begin{equation}
\sum_j h_j = \sum_j \int_0^{V_j} I_j(v)dv
\end{equation}
where the sum is over the resistors, and the free variables are the voltages
at each node \cite{ref1}.

While we are mainly interested in the infinite lattice case,
realistically we must work with lattices of finite size.  While 
one generally approximates the infinite case by averaging over
many small lattices, the finite size fluctuations can still
be important.
                                        
To understand the effect finite size has on the critical 
current, take a binary model on a square lattice.  
Eq.\ \ref{eq:bound} gave us the bound
\begin{equation}
{ I_{c1} \le \frac{I_{ctot}}{N} \le pI_{c1} + (1-p)I_{c2} }
\end{equation}
for an infinite lattice.
However, for a finite lattice, this upper bound decreases.
For a $N \times N$ lattice of nodes, we have a
resistor lattice with $ M = N-1$ columns and $N$ resistors in
each column Fig. \ref{fig:red}.  We are interested in the fluctuations. The
occupation probabilities may be $p$ and $q=1-p$, but we are obviously
going to see large fluctuations from the mean in a smaller lattice.
So what we want is the expectation value for the maximum number
of $I_{c1}$ resistors in any of the $M$ columns, since this will
be the limiting factor on the critical current.  In order to
find this expectation value, we need the probability ${\cal P}(a)$ that
the maximum number of $I_{c1}$ resistors in any of the 
$M=N-1$ columns is $a$ (out of $N$).

To derive an expression for ${\cal P}(a)$, define $f(a)$ to be the probability
that a specific column has $'a'$ $I_{c1}$'s in it.  Then
\begin{equation}
{f(a) = \frac{N!}{a!(N-a)!}p^{a}(1-p)^{N-a} }.
\end{equation}
Now define $P_{M}(x,a)$ as the probability that x columns 
(out of $M$) have $a$ $I_{c1}$ resistors in them.
\begin{equation}
{P_{N}(x,a) = \frac{M}{x!(M-x)!} f(a)^{x} (1-f(a))^{M-x} }
\end{equation}
where $0 \le x \le M$.  ${\cal P}(a)$ is then given by:
\begin{eqnarray}
 {\cal P} (a) =& \sum _{x=1}^{M} [ P_{M}(x,a) \times \prod _{i=a+1}
^{N} P_{M-x}(0,i) ] & \qquad a \ne N \nonumber\\
  {\cal P} (N) =& \sum_{x=1}^{M} P_{M}(x,a) & \qquad a=N \nonumber\\
\end{eqnarray}
The above equation is simply the probability that x columns in M have
$a$ $R_{1}$ resistors, multiplied by the probability that there are  
no
columns (in the $M-x$ remaining) that have more than $a$ $R_{1}$ 
resistors in them, summed over x.

We then use ${\cal P}(a)$ to find the expectation value of a:
\begin{equation}
{ \bar a = \frac{\sum_{a=0}^{N} a {\cal P}(a)}{ \sum _{a=0}^{N}  
{\cal P}(a) }    }
\end{equation}
Finally we get the critical current for the lattice, 
\begin{equation}
I_{ctot} = \frac{\bar a I_{c1} + (N- \bar a ) I_{c2}}{N}
\end{equation}

\begin{figure}
\begin{center}
\leavevmode
\epsfxsize \columnwidth
\epsffile{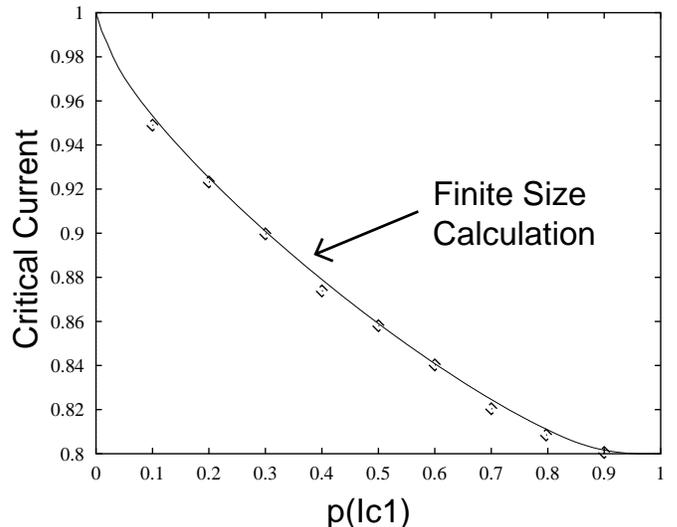}
\caption[]{Finite size calculation.  $I_{c1}/I_{c2} = 0.8$  
The limiting effects of finite size on the critical current are calculated
for a $15 \times 15$ square lattice and compared with the numerical results.}
\label{fig:finite}
\end{center}
\end{figure}

This analysis is only relevant for small lattices, since for 
very large lattices $\bar a$ will be very close to $p$ .   However, for 
smaller lattices this effect can indeed be the limiting factor, 
as can be seen
by looking at Fig.\ \ref{fig:finite} 
where we plot the critical current for a 15 by 15 lattice
with $I_{c1}=0.8$ and $I_{c2}=1.0$ for various occupation probabilities.
The critical current is limited by the 
finite size of the lattice and that the critical current for the infinite
case would be larger. 

\section{$I-V$ Relations}

The current-voltage characteristics of both JJ and FF binary models 
were studied.  For the sake of computational ease,  the
calculations were performed on square lattices, usually 10 by 10.
The $I-V$ characteristics from 50 lattices 
were averaged to obtain the final result.
It should be noted that the JJ $I-V$ characteristic
used in the calculations is slightly modified from that which was 
previously discussed. 
Instead of a strictly discontinuous model, the transition at $I_c$ from
$V = 0$ to $V = IR$ was made to be linear 
over a small current range $\delta I$.
In addition to being more physically realistic, this modification
made the convergence of the simulated annealing algorithm much quicker. 

\begin{figure}
\begin{center}
\leavevmode
\epsfxsize \columnwidth
\epsffile{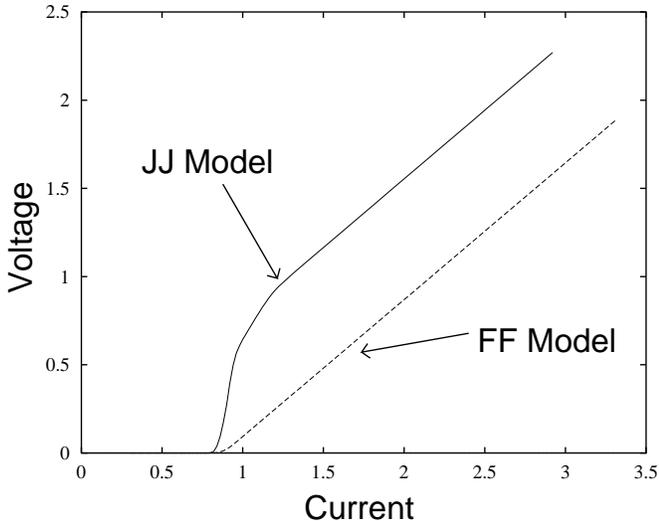}
\caption[]{Comparison of the $I-V$ characteristics for binary
 JJ and FF models on a 10 by 10 percolative lattice.  Both models possess the 
 same distribution of microscopic critical currents
 and normal state resistances.  ($I_{c1}/I_{c2}=0.75$ ,
$R_1/R_2 = 0.75$ ,$p=0.5$)  The {\it shape} of the individual circuit elements
$I-V$ characteristics is preserved when they are combined into a lattice.}
\label{fig:JJvsFF}
\end{center}
\end{figure}

The most important result of the numerical simulations is that the 
overall shape of the individual elements V-I characteristics is
preserved when the elements are combined into a lattice.  Fig.\ \ref{fig:JJvsFF}
compares a 10 by 10 lattice composed of JJ resistors with one of FF. 
$p=0.5,$  $I_{c1}/I_{c2} = 0.75$
and $R_1 / R_2 =0.75 $ in both cases.  We see that the resulting overall 
characteristics are very different.  This in itself is not surprising
as the individual components have different properties.  However,
the fact that the overall characteristics should resemble that of the individual 
elements so closely is surprising.  Statistical averaging, even with strong
nonlinearity, does not wash out the underlying input.

Since both JJ and FF resistors are linear at high
current, it is clear that the collective behavior of a lattice of resistors
will be linear at high currents.  In addition, the slope (conductivity)
will be independent of
the model used.  We also expect that this high current behavior of the 
JJ and FF V-I curves 
is  offset vertically, so that for JJ we expect the voltage to go as
${\cal V = I R }$ and for FF, $ {\cal V = ( I - I_{\rm c}^{\rm eff} ) R }$. 
(${\cal I_{\rm c}^{\rm eff}} \ne I_{ctot} $ necessarily )
This can be seen in Fig.\ \ref{fig:JJvsFF}.  The nature of the 
transitions between the superconducting and the high current behavior
as well as the current range over which the transition should
occur is not as evident.

Our calculations show that the net transition behavior closely mimics
that of the individual elements.  Qualitaively, we believe
the explanation is as follows.  
While $I_{ctot}$ is independent of the choice of model, as there
is no difference in the behaviour of JJ and FF resistors below $I_c$,
the behaviour above $I_{ctot}$ is highly model dependent.
FF resistors can conduct at all
voltages, and hence the transition from superconductivity to dissipative
conduction for a FF lattice is smooth and gradual.  However, the JJ resistors have 
an almost discontinuous jump in resistance at $I_c$.  It is harder to 
satisfy Kirchoff's laws at low voltages, since these low voltages are
in effect "forbidden".  Thus the rise in voltage 
at $I_{ctot}$ is very steep, and the breadth of the transition rather narrow.  

\begin{figure}[t]
\begin{center}
\leavevmode
\epsfxsize \columnwidth
\epsffile{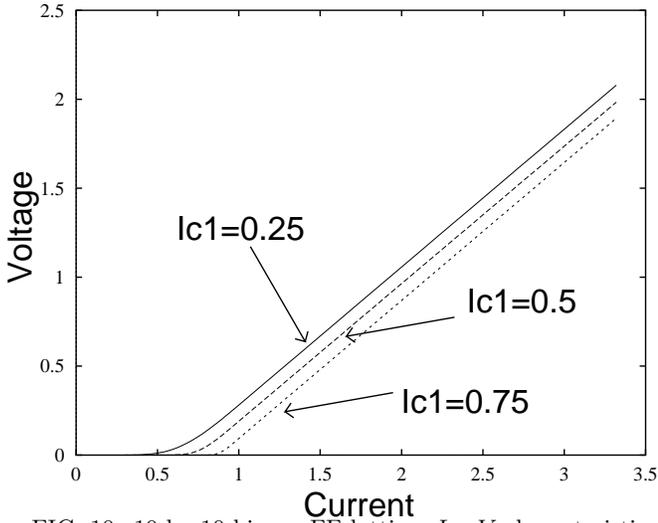}
\caption[]{10 by 10 binary FF lattice.  $I-V$ characteristics for various 
 $I_{c1}/I_{c2}$ are shown. $p(I_{c1})=0.5$ , $R_1/R_2 =0.75$ }
\label{fig:fig10}
\end{center}
\end{figure}

\begin{figure}[t]
\begin{center}
\leavevmode
\epsfxsize \columnwidth
\epsffile{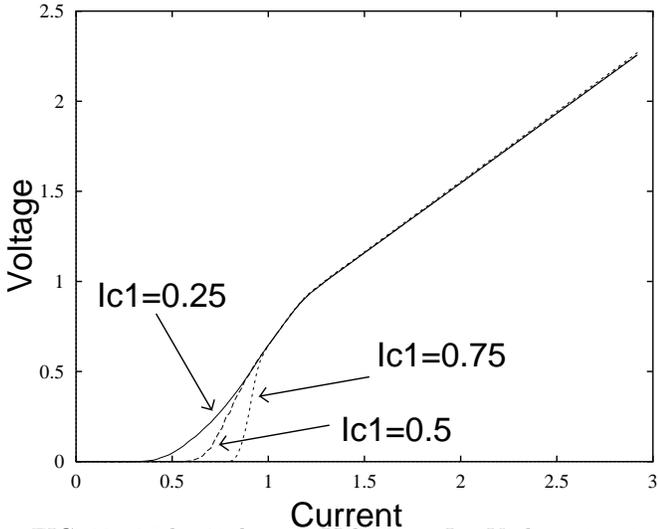}
\caption[]{10 by 10 binary JJ lattice.  $I-V$ characteristic for various
 $I_{c1}/I_{c2}$ are shown.  $ p(I_{c1})=0.5$ , $R_1/R_2=0.75$ \ \ There is 
 substantial broadening of the transition near $I_c$ due to the finite
 lattice size.}
\label{fig:fig11}
\end{center}
\end{figure}

\begin{figure}
\begin{center}
\leavevmode
\epsfxsize \columnwidth
\epsffile{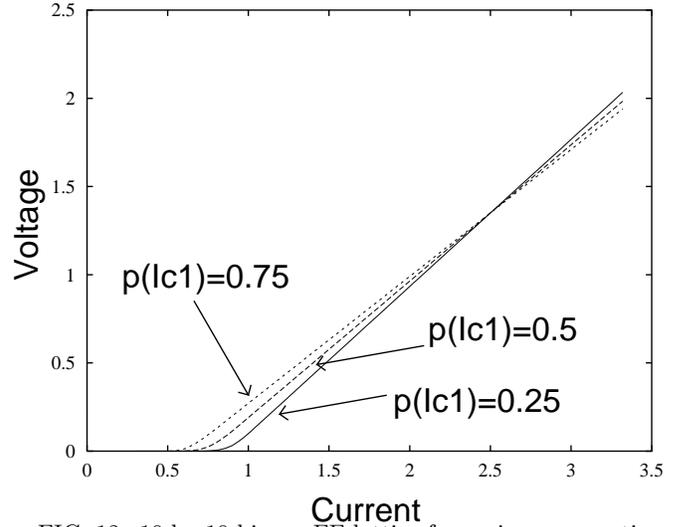}
\caption[]{10 by 10 binary FF lattice for various occupation probabilities
$p(I_{c1})$. $I_{c1}/I_{c2} = 0.5$ , $R_1/R_2 = 0.75$}
\label{fig:fig12}
\end{center}
\end{figure}

\begin{figure}
\begin{center}
\leavevmode
\epsfxsize \columnwidth
\epsffile{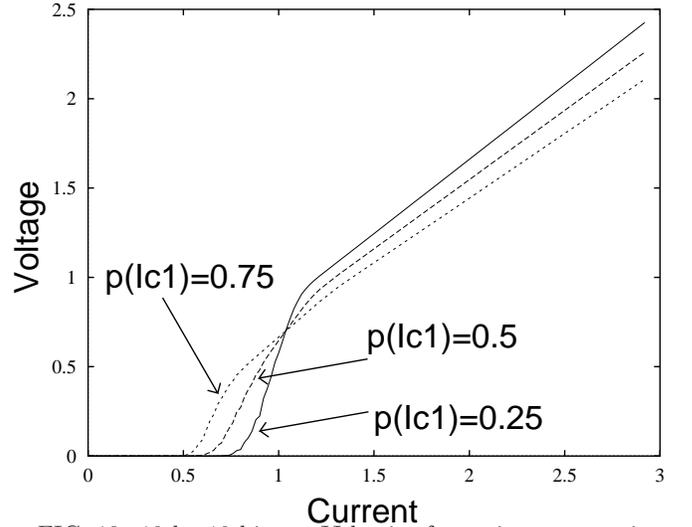}
\caption[]{10 by 10 binary JJ lattice for various occupation probabilities
$p(I_{c1})$. $I_{c1}/I_{c2} = 0.5$ , $R_1/R_2 = 0.75$  Again, finite
size transition broadening may be seen.}
\label{fig:fig13}
\end{center}
\end{figure}

Exactly how narrow the transition to dissipative behaviour is depends on the 
probability distribution of the critical currents.  
For the binary model, as $I_{c2} - I_{c1}$ grows, the transition broadens.
This is true for both JJ and FF, (although it is easiest to see in the JJ case). 
We illustrate this in Figs.\ \ref{fig:fig10} and \ref{fig:fig11} where the
effects of varying $I_{c1}/I_{c2}$ for lattices of fixed occupation probability
$p (q=1-p)$ and fixed $R_1 / R_2 $ for both JJ and FF lattices are shown.
If on the other hand we vary the 
occupation probability $p$ as in Figs.\ \ref{fig:fig12} and 
\ref{fig:fig13}, we see that the transition from
superconductivity to the high current behavior is broadest at $p=0.5$,
and gets much narrower towards $p=0$ or 1.  Thus we see that as the
{\it probability distribution} for the critical currents of the individual
elements flattens out, the current range occupied by the transition from
superconductor, to normal conductor becomes much larger.  This statement
is still relevant for non-binary systems which have a distribution of
$I-V$ characteristics ${\cal P}(V(I))$.

Very detailed comparison between theory
and experiment is beyond the scope of this 
paper.  It would 
require the experimental
determination of the entire individual resistor distribution
as well as macroscopic $I-V$ curves.
However, we can get partial information about the former 
and compare to the latter as a preliminary exercise.

Fig.\ \ref{fig:tendeg} shows  $I-V$
characteristics measured across a single $10^{\circ}$ grain boundary
at various applied fields. 
Varying the magnetic field applied to a grain boundary will change the
grain boundary's properties.  The critical current, in general, decreases
with increased applied field.  The high current resistance is dependent
on the applied field as well.  We however wish to focus for now on the 
{\it shape} of the boundary's $I-V$ characteristic.
At low
fields the grain boundary
displays a Josephson junction type behaviour which
resembles our JJ model.  At about 6 T there is a transformation to
a different high field behaviour caused by the depinning of flux along
the grain boundary.  We modeled this high field behavior with our FF model.  

How does a sample consisting of many grain boundaries behave?
Fig.\ \ref{fig:ibad} gives the voltage current characteristics for
an link comprised of many grain boundaries, made by 
ion-beam assisted deposition (IBAD).  At low
applied fields, the overall $I-V$ characterisic looks somewhat like a 
Josephson junction.  As the field is increased
the curvature of the characteristic changes at about 5 T, and 
it begins to look much more linear.  It appears that
the new dissipative behavior is dominated by flux creep as opposed
to a Josephson type dissipation.  Our resistor model supports this.
We have shown through our resistor model that the microscopic 
single grain boundary (resistor)
characteristics can (at least in the case of JJ and FF) be carried 
through by percolation to the macroscopic many grain boundary case.  
A macroscopic sample comprised of grain boundaries acting as 
Josephson junctions will itself have a very sharply rising 
$I-V$ characteristic, while
a sample where the dissipation is dominated by flux creep
will resemble its individual components as well.

Comparison of our theory with the experiment suggests that the grain boundaries 
dominate the dissipative properties
of the macroscopic superconductor, {\it even in applied magnetic fields}.
If this is the case, then the effects of the grains themselves 
are less important for the IBAD samples.
\section{Conclusion}

Conduction in polycrystalline high-T$_c$ superconductors is
percolative, each percolative path being nonlinear.  We believe that
the model considered in this paper is general enough to capture the
essence of this phenomenon.  It requires powerful numerical techniques:
the one introduced in this paper should be flexible enough to handle
virtually any distribution of grain boundary conductances.  For simple
distributions, for example binary models, probabilistic reasoning can 
give insight into, and even good quantitative approximations for,
the magnitude of the macroscopic critical current.  

Not only the critical current is of interest for applications.  
One also wishes to know what the full $I-V$ characteristic 
is telling us about the properties of individual grain boundaries.  
What emerges from the calculations is that the {\it shape} of macroscopic
current-voltage relation mirrors, to a remarkable extent, 
the {\it shape} of the underlying current-voltage relations
of the grain boundaries.  This is counterintuitive because the
latter have a distribution which may be broad, and one would expect
this to produce a broadening which would wash out discontinuous
behavior near $I_{ctot}$.  This does not happen.  Nor does it happen that,
for example, there is a two-step behavior for binary distributions.
Instead, we find that the full $I-V$ always has nearly the same 
shape as the individual $I-V$'s, but with parameters which are averages
over the underlying distribution.

Our method allows us to find the full spatial distribution of
current.  We have not yet made use of this information.  It is
measurable by magneto-optical or Hall probe methods, 
and surely contains information about the spatial distribution of
microscopic critical currents.  This represents a possible future direction
for this work.

We would like to thank A. Gurevich, E. Pashitskii, M. Friesen,
N. Heinig, D. Cyr
and D.C. Larbalestier for useful discussions.
This work is supported by the NSF under 
Grant No. DMR-9704972 and under the
Materials Research Science and Engineering Center Program,
Grant No. DMR-96-32527.

\end{document}